\documentclass{optica-article}

\journal{opticajournal} 

\articletype{Research Article}

\usepackage{lineno}
\usepackage{soul}

\begin{document}

\title{Noise analysis of a quasi-phase-matched quantum frequency converter and higher-order counter-propagating SPDC}

\author{Felix Mann\authormark{1,*}, Helen M. Chrzanowski\authormark{1}, Felipe Gewers\authormark{1,2}, Marlon Placke\authormark{1} and Sven Ramelow\authormark{1,3}}

\address
{
\authormark{1}Institut f\"ur Physik, Humboldt-Universit\"at zu Berlin, Newtonstr. 15, 12489 Berlin, Germany\\
\authormark{2}Instituto de Física, Universidade de São Paulo, R. do Matão 1371, São Paulo 05315-970, Brazil\\
\authormark{3}IRIS Adlershof, Humboldt-Universität zu Berlin, Berlin, Germany
}

\email{\authormark{*}felixmann@physik.hu-berlin.de} 




\begin{abstract}
Quantum frequency conversion (QFC) will be an indispensable ingredient in future quantum technologies. For example, large-scale fibre-based quantum networks will require QFC to interconnect heterogeneous building blocks like emitters, channels, memories and detectors. The performance of existing QFC devices - typically realised in periodically-poled nonlinear crystals - is often severely limited by parasitic noise that arises when the pump wavelength lies between the inter-converted wavelengths. Here we comprehensively investigate the noise spectrum of a QFC device pumped by a CW 1064 nm laser. The converter was realised as a bulk periodically-poled potassium titanyl phosphate (ppKTP) crystal quasi-phase-matched for conversion between 637 nm and 1587 nm, which was also polished and coated to resonantly enhance the pump field by a factor of 50. While Raman scattering dominates the noise background from 1140 nm to 1330 nm, at larger energy shifts (beyond 60 THz), parasitic spontaneous parametric down-conversion (SPDC) is the strongest noise source. Further, the noise spectrum was contaminated by a regular succession of narrow-band peaks, which we attribute to a heretofore unidentified higher-order counter-propagating SPDC processes - with quasi-phase-matching orders up to 44 evident in our measurements. This work provides a comprehensive overview of the limiting noise sources in QFC devices that use quasi-phase-matched crystals and will prove an invaluable resource in guiding their future development.
\end{abstract}

\section{Introduction}
The efficient and noiseless frequency conversion of quantum light will be an essential building block in the development of future large-scale quantum networks. By enabling the coherent transfer of quantum information between disparate energies, it interfaces the low-loss capabilities of telecommunications infrastructure with the typically visible and near-IR photon energies required by matter-based qubits. One such example are nitrogen-vacancy (NV) centres in diamond ~\cite{hansonNV,hanson_color,munroNV}. NV centres represent a promising platform due to the long coherence time of the electronic spin associated with this defect centre \cite{nv_coherence}, the possibility to couple this electronic spin to nuclear spin \cite{register,nv_memory,nv_memory2} and an optical interface at 637 nm \cite{wrachtrup}. With NV centres in diamond, quite complex quantum networking tasks have been recently demonstrated~\cite{hanson_networking1, hanson_networking2, hanson_networking3}. Without the conversion of the 637 nm photons emitted by the NV centre to telecommunication wavelengths, these photons would suffer dramatic transmission losses in optical fibres \cite{dreau} that limit the size of these networks well below the metropolitan scale. A variety of platforms are similarly limited by this, including other colour centres in diamond like GeV (602 nm), SnV (620 nm), PbV (520/555 nm)~\cite{hanson_color} or quantum memory at 606 nm~\cite{memory606} and 580 nm~\cite{memory580}, trapped ions (369.5 nm)~\cite{ions2} or hexagonal boron nitride (\raisebox{-0.9ex}{\~{ }}600 nm)~\cite{hBN_1}.

In these scenarios, quantum frequency conversion (QFC) is often realized as a difference- (sum-) frequency generation process in which a $\chi^{(2)}$-nonlinear crystal converts a typically short wavelength photon to the telecommunications band (or vice versa) via a comparatively strong pump field at an intermediate wavelength. In this configuration, where the pump wavelength lies between the emitter and target wavelengths, the strong field required to ensure near-unity conversion efficiency also produces considerable unwanted noise at the target wavelength. This parasitic noise is not only present in the conversion from visible to telecommunication wavelengths, but also for the reverse process, since, for an efficient converter, any noise produced will itself be efficiently up- or down-converted. This parasitic noise -- whether parametric or non-parametric in origin -- has proven to be the limiting factor in the performance of these conversion devices.

Efficient conversion demands a strength of non-linear interaction that is typically inaccessible in a three-wave-mixing process without either waveguide confinement~\cite{dreau}, pump enhancement~\cite{albota} or pulsed pump lasers~\cite{observation}. Furthermore, frequency converters near-exclusively utilise quasi-phase-matching, as it simultaneously allows for collinear interactions of the fields without spatial walk-off and unlocks access to the largest tensor elements of the nonlinear susceptibility. In theory, this allows for a dramatic reduction in pump power that ideally mitigates noise from non-parametric scattering of the pump photons, notably, Raman scattering and fluorescence. However, errors that arise in the periodic poling can vastly enhance the otherwise strongly suppressed phase-mismatched parasitic spontaneous parametric down-conversion (SPDC) processes. Furthermore, employing waveguide confinement to reduce the needed pump power can lead to a Cerenkov-idler configuration of the SPDC noise process \cite{rastogi} and waveguide inhomogeneities to further undesired phase-matching, both increasing the noise level even more.

This emergent noise floor in the conversion between visible and telecommunication wavelengths was first reported by Albota and Wong \cite{albota} who considered conversion in a bulk periodically-poled lithium niobate (ppLN) crystal. Almost two decades on, converters based on ppLN waveguides, which remain the most commonplace platform, are still largely limited by pump-induced noise~\cite{pelc1,ikuta,rutz,maring,afzelius,dreau}. The noise spectra of quantum frequency converters based on such devices have been previously investigated in references \cite{pelc_raman,afzelius}. In the short-wavelength-pumped scheme of Ref. \cite{afzelius}, parasitic SPDC is found to be the dominant source of noise. In the long-wavelength-pumped converter investigated in Ref. \cite{pelc_raman}, anti-Stokes Raman scattering is identified as the main source of noise.

A workaround for this noise problem is two-step conversion. Here the pump wavelength is chosen to be longer than the target wavelength (e.g. 2.1 $\mu m$) and, due to energy conservation, no unwanted SPDC can be generated at the target wavelength. Additionally, the Raman shift for the anti-Stokes photon is very large, ensuring its noise contribution is negligible. First proposed in \cite{albota}, two-step conversion was first demonstrated in a single waveguide device \cite{twostep} and later in two separate waveguides~\cite{becher}. A drawback of this approach is the necessity for a strong laser source at about 2 $\mu m$ alongside the increased complexity of a two-step process. Another promising approach to circumvent the parasitic noise arising in conversion is to utilise critical phase-matching in a bulk crystal~\cite{geus}, eliminating parasitic SPDC noise which would be enhanced by poling errors. Owing to the comparatively smaller nonlinear susceptibility and the effects of double refraction, this approach demands substantially more pump power and the maximum conversion efficiency is limited by the spatial walk-off encountered in critical phase-matching. By utilising external resonant enhancement of the pump, this approach achieved an internal conversion efficiency of 60\% for a circulating pump power of 360~W while obtaining a low measured noise of 2.0 kHz/nm. Thus, this approach has about a factor of 20 less noise than a bulk ppKTP converter \cite{mann2} and two-orders of magnitude less noise then the best ppLN waveguide \cite{dreau}.

In our previous work, we demonstrated an alternative route for low-noise QFC based on a single-step quantum frequency converter based on a monolithic cavity made from bulk periodically-poled potassium titanyl phosphate (ppKTP)~\cite{mann2}. The converter was designed for the conversion of single photons at a wavelength of 637 nm to the target wavelength in the telecommunication L-band (1587 nm) with a strong 1064 nm CW pump laser. Due to the improved poling quality of ppKTP~\cite{canalias,mann1}, parasitic noise arising from spurious SPDC was reduced by a factor of five compared to the best ppLN waveguide~\cite{dreau}, while attaining a high internal conversion efficiency of 72\% at about 72 W circulating pump power. The converter, however, still exhibited a pump-induced noise floor of 110 kHz/nm at the target wavelength. Considerable improvements in the noise floor are required for single-step converters based on quasi-phase-matched nonlinear crystals to be a viable technology for frequency conversion between the visible and telecommunication wavelengths. 

\begin{figure}[!htbp]
\centering
\includegraphics[width=11cm]{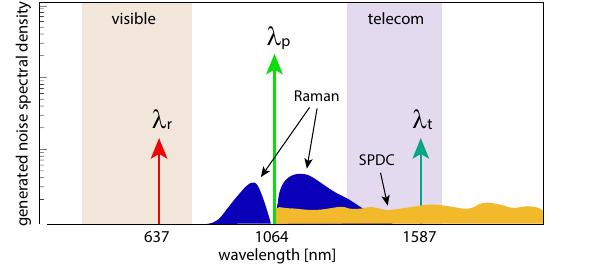}
\caption{Sketch of the conversion scheme between visible and telecommunication wavelengths with the strong pump mode in between, outlining potential noise sources. The pump photons can undergo Raman scattering to longer and, to a lesser extent, to shorter wavelengths. Pump photons can also undergo parasitic SPDC, producing a photon pair at longer wavelengths, with a signal photon possessing an energy closer to that of the pump. In converters realised using periodically-poled crystals (whether a bulk or waveguide implementation) an SPDC noise pedestal arises due to random duty cycle errors in the poling. The noise spectral density (NSD) on the y-axis represent the noise count rates per spectral bandwidth.}
\label{fig:sketch}
\end{figure}
In this work, we investigate the noise spectrum of the aforementioned bulk ppKTP quantum frequency converter designed to convert photons between 637 nm and 1587 nm via a strong 1064 nm pump laser. Our measurements fully characterise the resulting noise spectrum in the wavelength range from 1140 nm to 1650 nm. By resolving the polarisation and temperature dependence of features within the noise spectrum, we identify different noise contributions and the wavelength regions where they dominate. These results are of immediate relevance for most schemes where the strong pump field lies between the two interconnected wavelengths. 

Fig.~\ref{fig:sketch} provides a sketch of our conversion scheme and the primary noise contributions from the different processes. We find that close to this pump wavelength, Raman scattering of pump light is the dominant noise source. As we consider increasingly longer wavelengths approaching the target wavelength, the required phonon energies increase and the contribution of Raman scattering to the noise background is quickly suppressed and spurious spontaneous parametric down-conversion (SPDC) becomes the dominant noise process. This is near-exclusively the case for quasi-phase-matched crystals, where errors in the periodic poling~\cite{fejer} drive spurious SPDC processes that produce a plateau in the noise floor, effectively stable across the transparency of the material. This effect has been studied in detail in \cite{pelc1,pelc2,phillips}.

In addition to the anticipated features from Raman scattering and the parasitic SPDC background, we have also identified an exotic nonlinear process - higher-order counter-propagating SPDC - that to our knowledge has never before been experimentally observed. This process, which is only present in periodically-poled devices, can contribute significantly to the parasitic noise background. Notably, both types of counter-propagation, where either the idler or signal photon counter-propagate relative to the pump beam, are observed, with narrow-band peaks of corresponding quasi-phase-matching orders m from 10 up to 44 visible. Owing to poling errors related to the duty cycle, peaks associated with even orders of m are also visible.

In totality, these results provide a comprehensive overview of the noise spectrum of a quantum frequency converter that employs a quasi-phase-matched process and a pump that lies in between the interconnected wavelengths. Our presented measurement techniques and findings offer the possibility to better understand and design periodically-poled nonlinear devices for photonic quantum technologies, e.g. quantum frequency converters for quantum networking.

\section{Experimental setup}
\begin{figure}[!htbp]
\centering
\includegraphics[width=10cm]{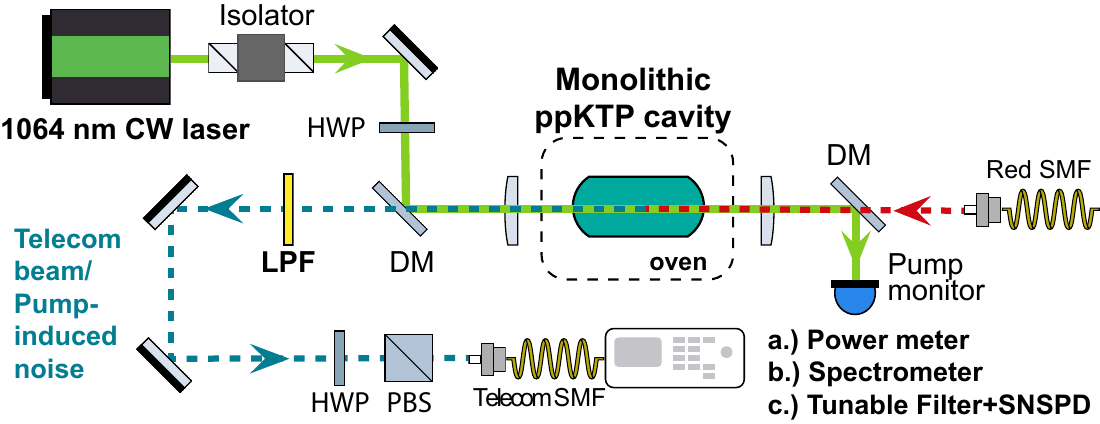}
\caption{Experimental setup for the noise analysis of the frequency converter based on the monolithic bulk ppKTP cavity: A CW 1064 nm ND:YAG pump laser with 3 W of optical power is coupled into the monolithic cavity. {\bf a.)} To define the correct output spatial-mode for the subsequent noise measurements, the conversion efficiency of the frequency converter is first optimised. For conversion, red light at 637 nm is launched from a single mode fibre (SMF) backwards into the ppKTP crystal, where it undergoes difference frequency generation with the circulating pump light, generating light at the target telecommunication wavelength. The beam is cleaned from pump light via a dichroic mirror (DM) and a long-pass filter (LPF). The polarisation of the detected light is selected via a half-wave plate (HWP) and a polarising beam-splitter (PBS) and then coupled into a SMF. {\bf b.)} The noise spectrum is measured via two InGaAs single-photon spectrometers with ranges from 1140 nm to 1650 nm and 1320 nm to 1490 nm. {\bf c.)} High-resolution measurements over a reduced range of the spectrum are performed with a tunable filter and a superconducting nanowire single-photon detector (SNSPD).}
\label{fig:setup}
\end{figure}
The experimental setup for the noise analysis of the frequency converter is shown in Fig.~\ref{fig:setup}. The main element of the experiment is the converter: a bulk ppKTP crystal which is polished and coated to also provide a monolithic enhancement cavity for the pump field~\cite{mann2}. We utilise a $L=20 \mbox{ mm}$ long flux-grown KTP crystal (from Raicol Crystals Ltd.) quasi-phase-matched for type-0 DFG (SFG) process 637 nm $\rightarrow$ 1064 nm + 1587 nm (or vice versa) with a poling domain length of $l = 7.85\mbox{ }\mu\mbox{m}$, making use of the strong $d_{33}$ tensor element of the nonlinear susceptibility for the parametric interaction. The conversion bandwidth is 110 GHz (FWHM), corresponding to 0.9 nm (0.15 nm) at 1587 nm (637 nm). The converter is pumped by a commercial 3W CW 1064 nm ND:YAG laser. For practical purposes - ensuring increased stability of the converter - all measurements presented here were performed at a circulating power of 50 W which yields an internal conversion efficiency $\eta_c \approx 54\%$. The circulating pump power was inferred from the transmitted pump power and the measured finesse $\mathcal{F}\approx 146$. The monolithic cavity is placed in an oven which is stabilised to less than $\pm1$ mK. Further details on the setup and monolithic ppKTP cavity are provided in \cite{mann2}.

A half-waveplate (HWP) enabled measurements for both linear pump polarisations, either aligned with the $z$- or $y$-axis of the crystal. The two polarisations have distinct resonance conditions, owing to the birefringence of the monolithic KTP resonator. Additionally, a HWP and a polarising beam-splitter (PBS) before the single mode fibre (SMF) defined the polarisation and the collection mode of the measured light. To ensure that the collection mode for our noise measurements matched the optimal output mode of the converter, the conversion was first optimised via the conversion of a weak probe at 637 nm. The 637 nm light was subsequently turned off and any residual pump laser light was filtered out with two long-pass filters.

The spectra presented in this work were taken with three different spectrometers: two 512 pixel InGaAs spectrometers (OceanOptics NIRQUEST+1.7) with spectral ranges of 900 nm to 1650 nm (resolution $\approx$ 1.82 nm) and 1320 nm to 1490 nm (resolution $\approx$ 0.55 nm). Both have a quantum efficiency of about 0.1\% in the standard-gain-mode and about 1\% in the high-gain-mode. We also performed a high-resolution measurement over a restricted region of the spectra from 1480 nm to 1620 nm with a resolution of approximately 0.1 nm. This measurement combined a tunable filter (EXFO XTA-50 U) with a transmission of about $40\%$ and a superconducting nanowire single-photon detector (SNSPD) with a quantum efficiency of about 90\% and a dark count rate of 100 cps (from Quantum Opus LLC).

\section{Experimental results}
\subsection{Measurements of the full spectrum}
\begin{figure}[!htbp]
\centering
\includegraphics[width=12.8cm]{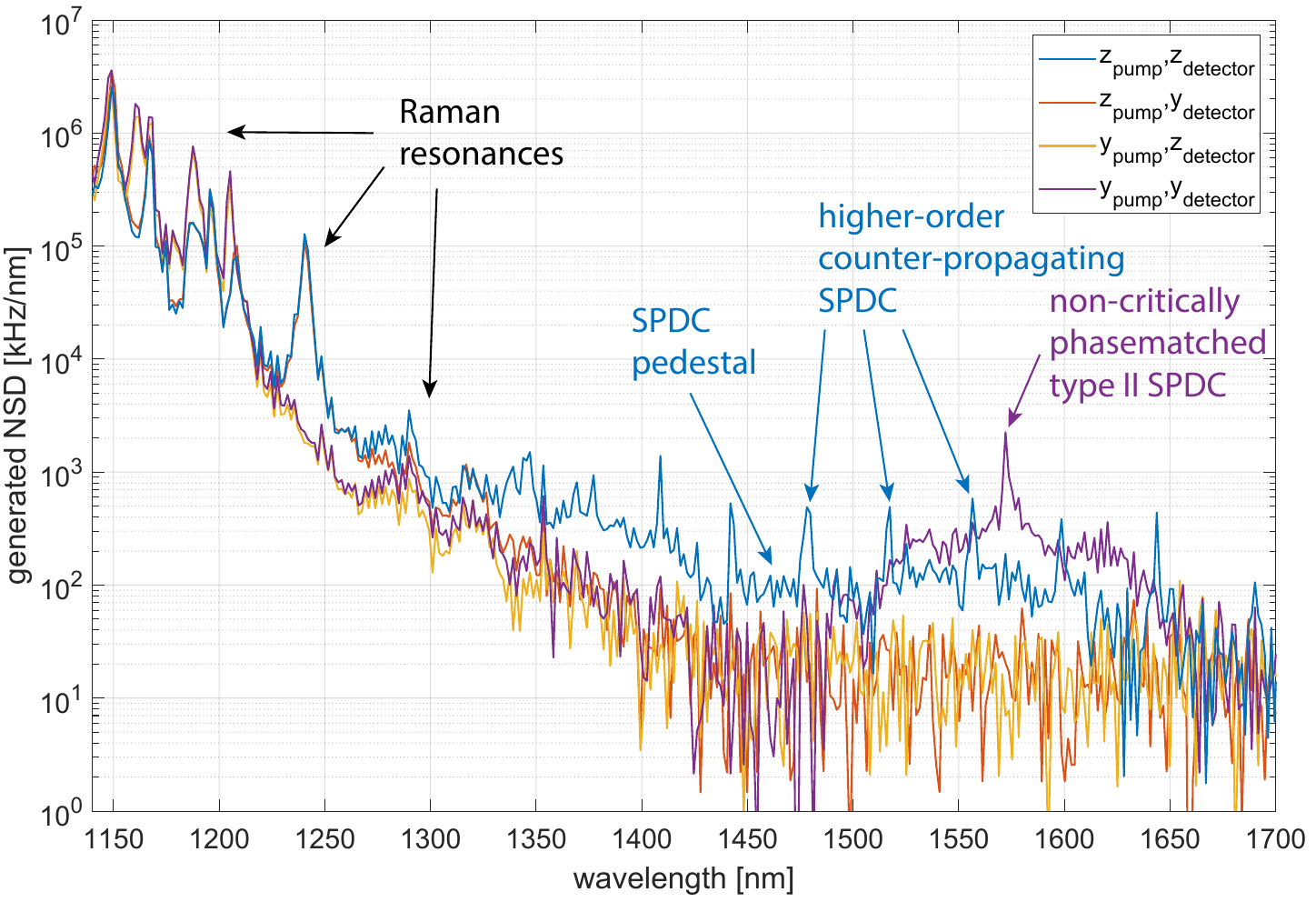}
\caption{Measured spectra from 1140 nm to 1650 nm for all four possible combinations for the polarisation of the pump field and the detected mode. Features associated with spontaneous Raman scattering, a broadband SPDC 'pedestal', and higher-order counter-propagating SPDC are all visible within the spectra. In addition, when the pump and detection polarisation are aligned along the $y-$axis of the crystal, a non-critically phase-matched type II SPDC process -- which is largely suppressed by the crystal grating -- is also visible. The spectra were recorded with an InGaAs single-photon spectrometer with a resolution of about 1.82 nm with an integration time of 30 seconds for each spectrum. The y-axis was normalised to the known generated noise spectral density (NSD) of 76 kHz/nm at the target wavelength of 1587 nm at 50 W circulating pump power~\cite{mann2}.}
\label{fig:all}
\end{figure}
Fig.~\ref{fig:all} shows the measurements with the wide-range InGaAs spectrometer from 1140 nm to 1650 nm with a resolution of about 1.82 nm. Below 1140 nm the LPF blocks the remaining pump light. Above 1650 nm the InGaAs line array is no longer sensitive. The four possible combinations of pump and detection polarisation are shown, with each aligned to either the $z-$ or $y-axis$ of the crystal. The blue line shows the measured spectrum for the polarisation configuration corresponding to the type 0 phase-matching condition (both pump and detected mode polarised along the $z$-axis of the crystal). This configuration is also the design configuration of the converter process. Due to the relative strength of the $d_{33}$ tensor element, this is the configuration where one would anticipate the largest noise floor in the region where parasitic SPDC is the dominant noise source.
\\The y-axis of Fig.~\ref{fig:all} is normalised to the known generated noise spectral density (NSD) of 76 kHz/nm at the target wavelength of 1587 nm at 50 W circulating pump power~\cite{mann2}. The crystal temperature was stabilized near $31.5\mbox{ }^{\circ}C$ with a long-term temperature stability of below 1 mK. All four spectra were corrected for the wavelength-dependent transmission of the optical coating of the monolithic cavity, which was specified by the manufacturer (Laseroptik GmbH).
\\In the range from 1140 nm to 1330 nm strong Lorenzian-shaped peaks that we attribute to spontaneous Raman scattering (see Section~\ref{chap:raman}) are visible. Beyond 1330 nm, the general trend of decreasing NSD is no longer evident, and an effective pedestal arises that we identify as parasitic SPDC (see Section~\ref{chap:parametric}). The apparent drop-off beyond 1650 nm is due to the insensitivity of the InGaAs sensor. The succession of narrow-band peaks in the blue graph are due to higher-order counter-propagating SPDC. High resolution measurements of these unique features are presented in Section~\ref{chap:parametric}. The theoretical background on higher-order counter-propagating SPDC is outlined in Section \ref{chap:counterpropagation}, which was mainly inspired by Fejer et al. \cite{fejer} and Phillips et al. \cite{phillips}.

The single peak evident near 1580 nm in the purple curve ($y_{pump},y_{detector}$) is attributed to a non-critically phase-matched type II process (1064 $\rightarrow$ 1571 nm + 3293 nm), which is heavily suppressed by the periodic poling of the crystal.

\subsection{Stokes-Raman scattering}
   \label{chap:raman}
\begin{figure}[!htbp]
\centering
\includegraphics[width=12cm]{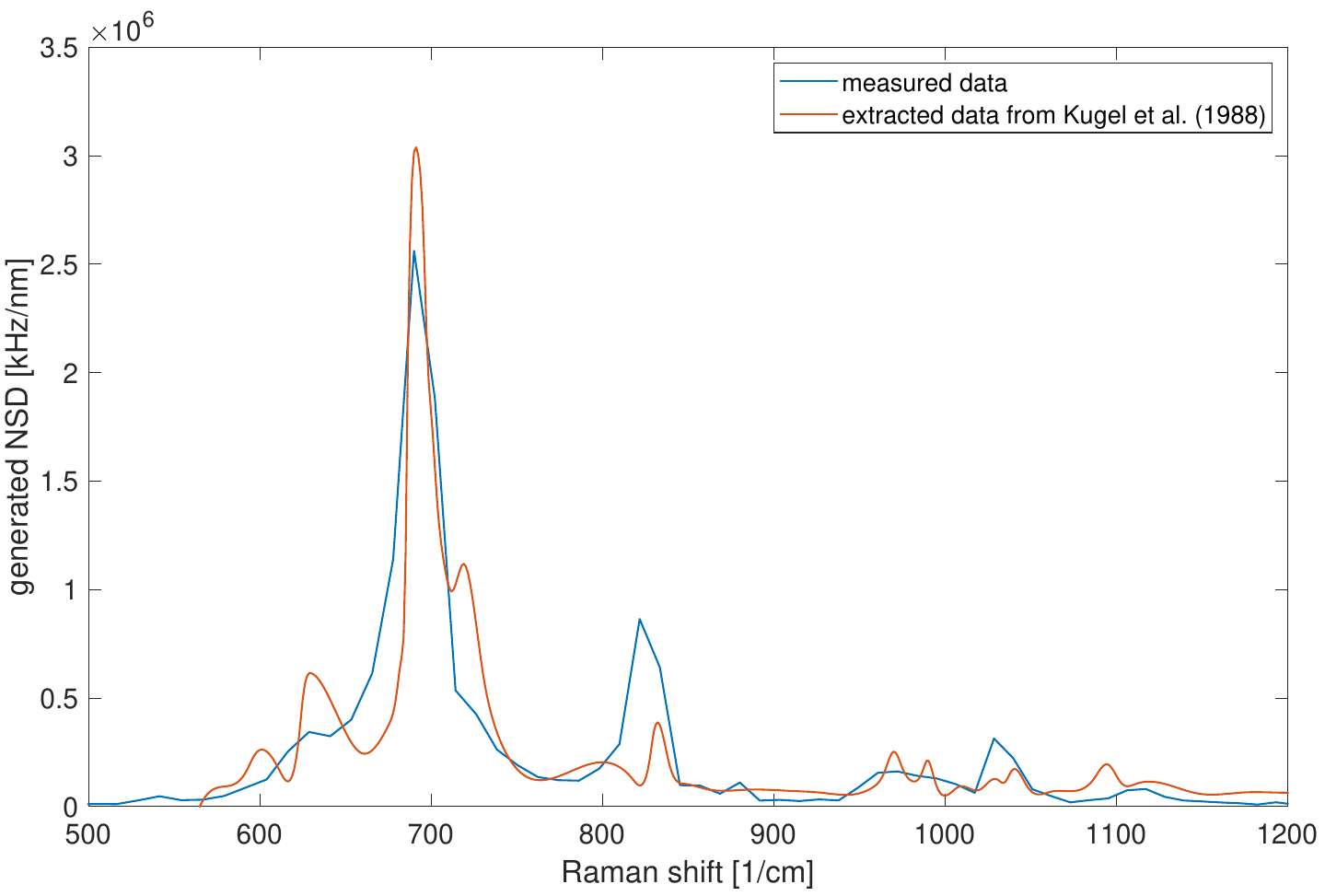}
\caption{Comparison of the measured spectrum ($z_{pump},z_{detector}$) to the results from Kugel et al. \cite{kugel}. The strongest peak was measured to be at $(692\pm 6)\mbox{ cm}^{-1}$ while in Ref. \cite{kugel} 687.5 $\mbox{cm}^{-1}$ was measured. This peak corresponds to the peak in Fig.~\ref{fig:all} at $(1149.4\pm 0.8)$ nm. See Table S1 in the supplemental document for a detailed comparison of the other peaks.}
\label{fig:raman}
\end{figure}
\begin{figure*}[!htbp]
\centering
\includegraphics[width=13cm]{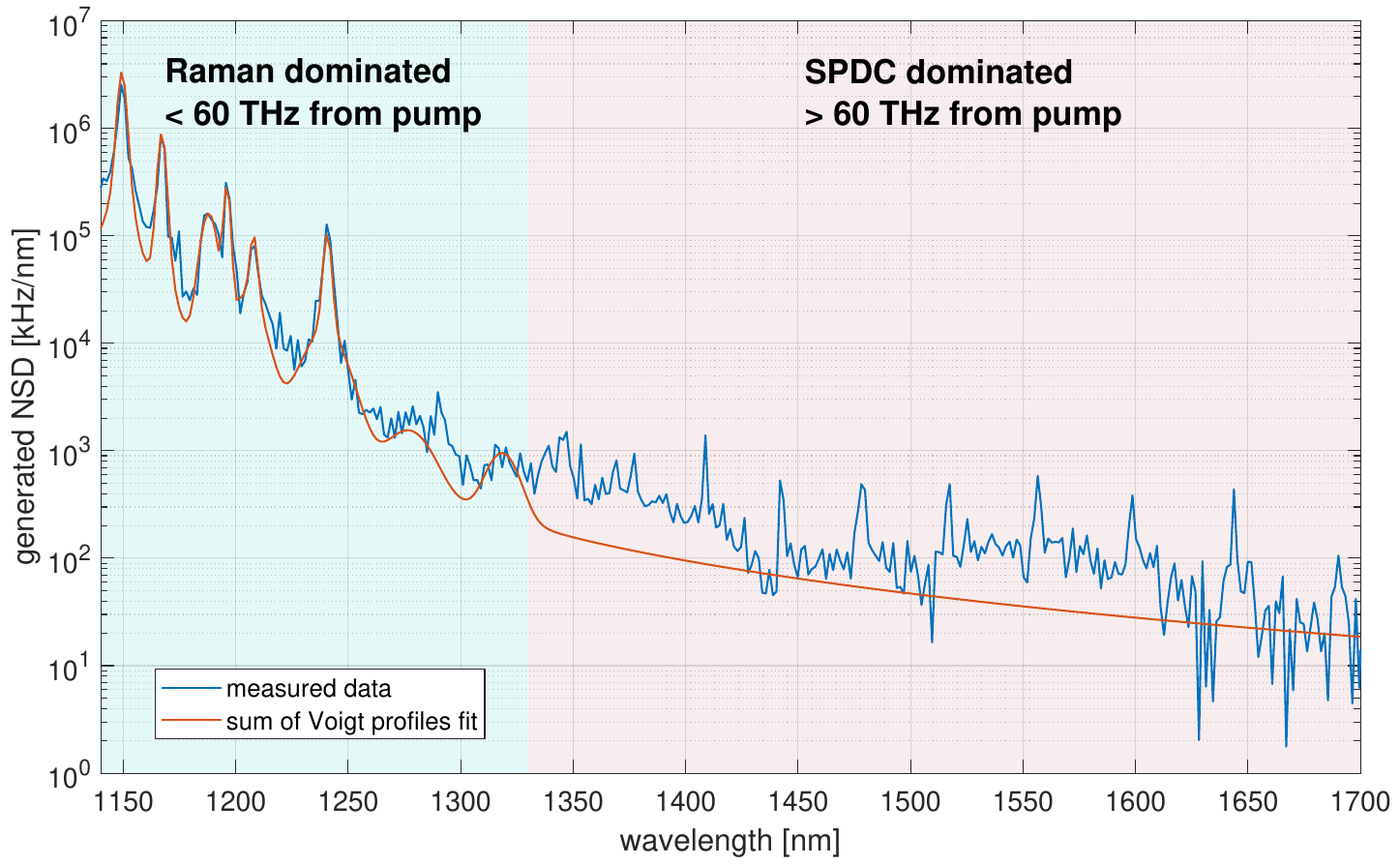}
\caption{Measured noise ($z_{pump},z_{detector}$) and a sum of Voigt profiles fit to ($z_{pump},y_{detector}$).}
\label{fig:tailing}
\end{figure*}

In Fig.~\ref{fig:raman} the blue curve ($z_{pump},z_{detector}$) from Fig.~\ref{fig:all} is plotted over the Raman shift from 500 $\mbox{cm}^{-1}$ to 1200 $\mbox{cm}^{-1}$. The Raman shift in terms of wavenumbers $\Delta \tilde{\nu}$ is calculated from the spectral wavelength $\lambda$ as $\Delta \tilde{\nu} = \left(1/\lambda_p-1/\lambda\right)$. The pump wavelength $\lambda_p$ was measured with a wavemeter to be ($1064.661 \pm 0.002$) nm. The resolution of the spectrometer over the plotted range is approximately 13 $\mbox{cm}^{-1}$. The orange curve is a Raman scattering measurement of unpoled KTP extracted from Fig.~5 in Ref. \cite{kugel} from the curve labelled with $A_1(X(YY)Z)$. The excitation wavelength in Ref. \cite{kugel} was 632.8 nm. See Table S1 in the supplemental document for a detailed comparison of the peak positions. Our measurements also qualitatively agree with the Raman measurements in Ref. \cite{ramanKTP1} for bulk KTP and for a proton-implanted KTP waveguide (compare Fig.~5 in Ref. \cite{ramanKTP1}) - both z-cut - where an excitation wavelength of 473 nm was used. Further, similar results were presented in Fig.~1 of Ref. \cite{ramanKTP2} with an excitation wavelength of 488 nm. As a result, we identify the strong peaks in the range from 1140 nm to 1250 nm (corresponding to a Raman shift of 621 $\mbox{cm}^{-1}$ to 1393 $\mbox{cm}^{-1}$) as Stokes-Raman scattering.
\\In Fig.~\ref{fig:tailing} the noise data was fitted with a sum of Voigt profiles following references \cite{voigt, faddeeva}. Using this approach, the contributions from the tail of the Raman scattering can be roughly disentangled from the contribution from parasitic SPDC. 

\subsection{Parametric noise and high-order counter-propagating SPDC}
   \label{chap:parametric}
\begin{figure*}[!htbp]
\centering
\includegraphics[width=11.7cm]{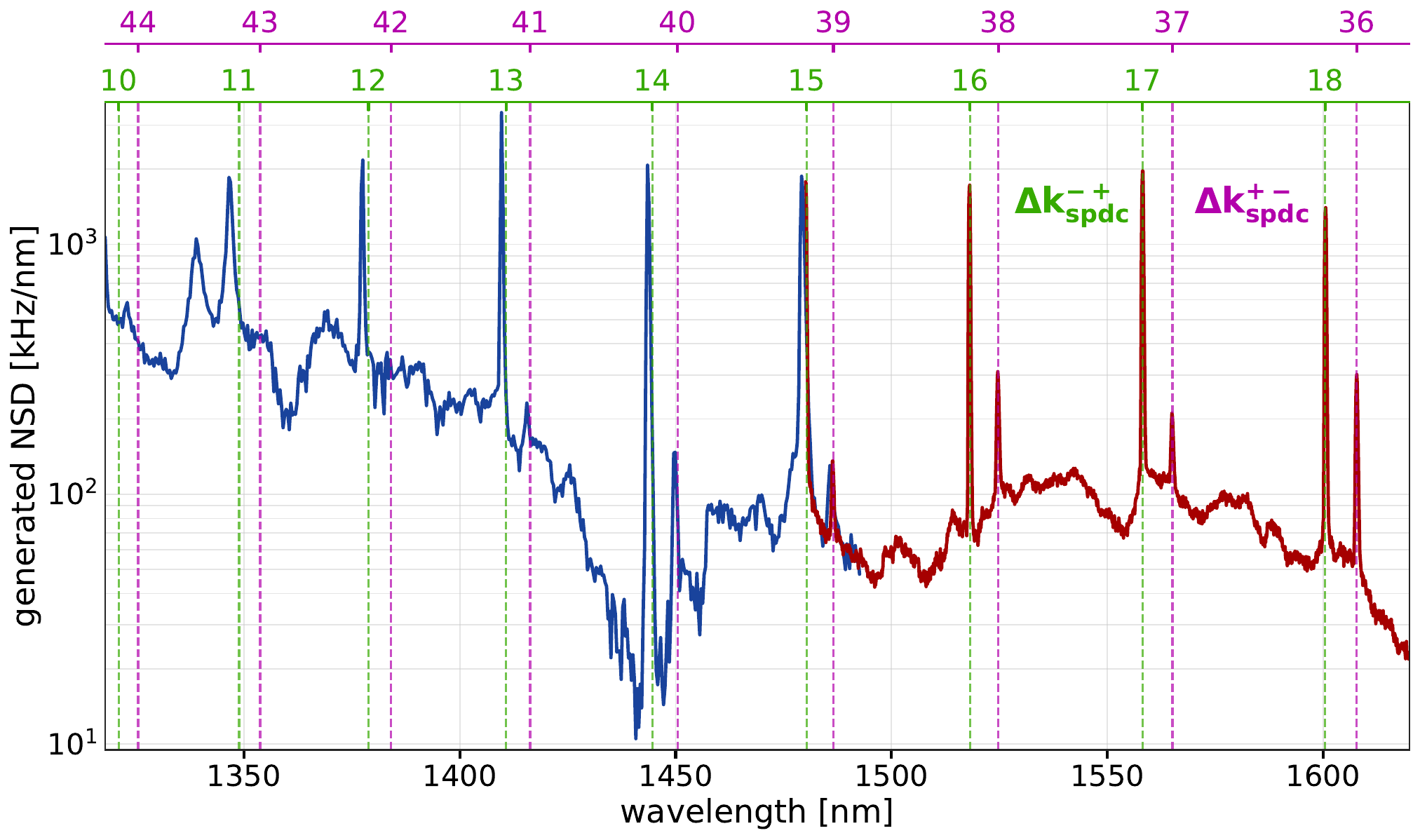}
\caption{High resolution measurements of the noise spectrum from 1320 nm to 1620 nm with the pump polarisation and detection aligned along the $z$-axis of the crystal. The presented spectrum was compiled with the combination of a narrow-band tunable filter and a SNSPD (red curve), and a InGaAs single-photon spectrometer (blue curve). A succession of sharp peaks of different heights from higher-order counter-propagating SPDC are observed emerging from a broadband, slowly varying SPDC background. The colored axes above the plot indicate the predicted order $m$ of the quasi-phase-matching, which is calculated from the ratio $\Delta k_{spdc}^{\mp \mp}/\Delta k_g$, where $\Delta k_g=2 \pi/ \Lambda$ is the grating vector and $\Lambda$ the poling period (compare Section 4 Equation 7). The "$\mp \mp$" indicates the direction in which the signal photons and the idler photons are travelling relative to the pump beam. Even-order higher-order peaks are present due to poling errors in the duty cycle. Towards shorter signal wavelengths (and longer idler wavelengths) the corresponding idler photons no longer lie in the transparency window of the crystal, with the inaccuracy of the Sellmeier equations in this region limiting the accuracy of the predicted phase-matching wavelengths. Likely due to material absorption of the idler reducing the effective crystal length, these narrow peaks also broaden considerably towards shorter signal wavelengths.}
\label{fig:counterpropagation1}
\end{figure*}

In this section we discuss the different contributions arising from SPDC in our noise measurements. Fig.~\ref{fig:counterpropagation1} presents a high resolution measurement of the noise spectrum from 1320 nm to 1620 nm. The red component of the curve indicates the region covered with a high-resolution measurement realised via the combination of a narrow-band tunable filter and a superconducting nanowire single-photon detector (SNSPD). The filter has a minimum bandwidth of 38 pm, a minimum step size of 1 pm, and a scan range from 1480 nm to 1620 nm. To maintain a feasible measurement time, the measurements were taken with both a bandwidth and step size of 100 pm. The blue portion of the graph was recorded with a InGaAs single-photon spectrometer from 1320 nm to 1490 nm (resolution $\approx$ 0.55 nm).
\\Fig.~\ref{fig:counterpropagation1} plots the measured noise background for the configuration where both the polarisation of pump and the detection are aligned with the $z$-axis of the crystal. The noise shows two distinct spectral feature: a slowly varying broad background and a succession of sharp peaks of different heights, which are not perfectly equidistant across the telecommunication wavelengths. We attribute the broad background noise to parasitic SPDC driven by poling errors and confirm this via temperature tuning of the crystal. The temperature tuning of spectral features of the broad noise floor roughly agrees with the prediction from temperature-dependent Sellmeier equations and the thermal expansion of the crystal\cite{kato,expansion}. See Fig.~\ref{fig:temptuning} and Table S2 in the supplemental document for more details on the temperature tuning measurements. For a circulating pump power of $(74.5\pm 0.3)$ W, we achieve our highest measured conversion efficiency of $(72.3\pm 0.4)$\% while generating $(110\pm 4)$ kHz/nm noise counts at the target wavelength of 1587 nm \cite{mann2}. This corresponds to a noise floor that is about 5 times smaller than that of the best state-of-the-art single-step converter which is based on a ppLN waveguide~\cite{dreau} and about 20 times larger than the unpoled bulk converter presented in Ref. \cite{geus}. Previous measurements of the phase-matching curve at very large detunings predicted a lower maximum for the noise floor of only 20 kHz/nm at unity conversion efficiency \cite{mann1}. We attribute this 5-fold discrepancy between the experimentally predicted and measured NSD to the difference in our theoretically anticipated and experimentally realised values of $P_{max}$. In the simple model for frequency conversion \cite{pelc1}, the NSD attributed to parasitic SPDC for unity conversion efficiency should be independent of both the pump power and the effective non-linearity. However, the converter efficiency deviated from its theoretically predicted dependence on pump power, with the experimentally inferred $P_{max}$ of 177 W being 2 to 3 times higher than that theoretically predicted. The measured NSD shows the expected linear dependence on pump power. One potential reason for the relatively large deviation between the predicted and inferred $P_{max}$ is a combination of effects not contained in the simple modelling, which does not fully incorporate imperfect mode-matching of the input beam, angle-anisotropy of the indices-of-refraction for the pump, input and output wavelengths and the spatially multi-mode nature of the input and output fields in the regime of high conversion efficiencies (input depletion) - the latter being an effect we observe in the mode-shape of the depleted input light after conversion at high efficiency.

Around the absorption edge of the crystal (idler wavelengths of 4.15 $\mu$m with a corresponding signal wavelength of 1430 nm) we observe an increase in the SPDC noise floor which then plateaus towards shorter signal (longer idler) wavelengths. One potential explanation for this increase in the noise floor is the partial absorption of the idler mode reducing the effective crystal length and therefore degrading the coherent suppression of highly phase-mismatched parasitic SPDC. In the literature the absorption edge of KTP is suggested to be between 4000 nm to 4500 nm \cite{ktp_transmission} and likely varies between different methods of crystal growth.

The series of narrow-band peaks are attributed to higher-order counter-propagating SPDC processes. These peaks emerge when the design poling period matches an integer multiple of the theoretical poling period for the counter-propagating process. As a result, these peaks lie equidistant in the wavevector mismatch $\Delta k_{spdc}^{\mp \mp}$. A detailed theoretical background on higher-order counter-propagating SPDC is given in Section~\ref{chap:counterpropagation}. Fig.~\ref{fig:counterpropagation1} plots the relevant measured spectra, with the order m of the processes denoted on the two axes above the plot for the two counter-propagating SPDC processes. The phase-matching order is predicted from the ratio $\Delta k_{spdc}^{\mp \mp}/\Delta k_g$, where $\Delta k_g=2 \pi/ \Lambda$ is the grating vector and $\Lambda$ the poling period. The "$\mp \mp$" indicates the direction in which the signal photons and the idler photons are travelling relative to the pump beam. Note that the quasi-phase-matching order for the counter-propagating signal (green) increases with increasing signal wavelength and vice-versa for the idler (pink). This is because the corresponding order necessary to satisfy phase-matching with the fixed grating will increase (decrease) with the increasing (decreasing) energy of the counter-propagating field. While for an idealised periodically-poled crystal with 50:50 duty cycle only odd orders of m are phase-matched, here the presence of poling errors in the duty cycle allows for phase-matching of the even orders. This is a consequence of the higher-order phase-matching, where an effective net contribution to the SPDC amplitude is only made in the last portion of each domain. Consequently, a deviation of only a few percent in the design duty cycle can amount to an appreciable deviation for a higher-order process where only the last few microns of each domain count.

Because we utilise a standing-wave cavity for the pump field, we collect signal photons corresponding to both counter-propagating processes in our measurements. For the process where the idler photon counter-propagates orders from $m=10$ to $m=18$ are visible in our measurements. For the alternate process, where the signal photon counter-propagates relative to the pump field, we observe orders from $m=36$ to $m=44$. As we move towards shorter signal wavelengths and accordingly longer idler wavelengths that lie beyond the transparent region of the crystal, we observe a broadening of the narrow peaks that we attribute to partial absorption of the idler field. This is especially evident for the 11th and 43th order peaks. Further, in this region where the crystal becomes increasingly opaque for the idler light, the accuracy of the Sellmeier equations \cite{kato} becomes limited and accordingly so to do our predictions for the exact phase-matching wavelengths.

To our knowledge, the observation of such higher-order counter-propagating peaks was first evident in the measurements of in Ref. \cite{mejia} in ppLN waveguides, though absent an explanation of their origin. In Ref. \cite{afzelius} the narrow-band peaks are not evident but may have been overwhelmed by the comparatively strong broadband noise. This is a noise source that should arise in all frequency converters based on quasi-phase-matched sources where the pump lies between the interconnected wavelengths. Notably, in the limit of an ideal quasi-phase-matched crystal - which would ideally suppress the contribution of parasitic phase-matching that plagues current implementations - these noise peaks are most pronounced - albeit only for odd orders. Owing however, to the narrow bandwidths of these features, they will most often lie outside the design bandwidth of a converter.
\begin{figure*}[!htbp]
\centering
\includegraphics[width=13cm]{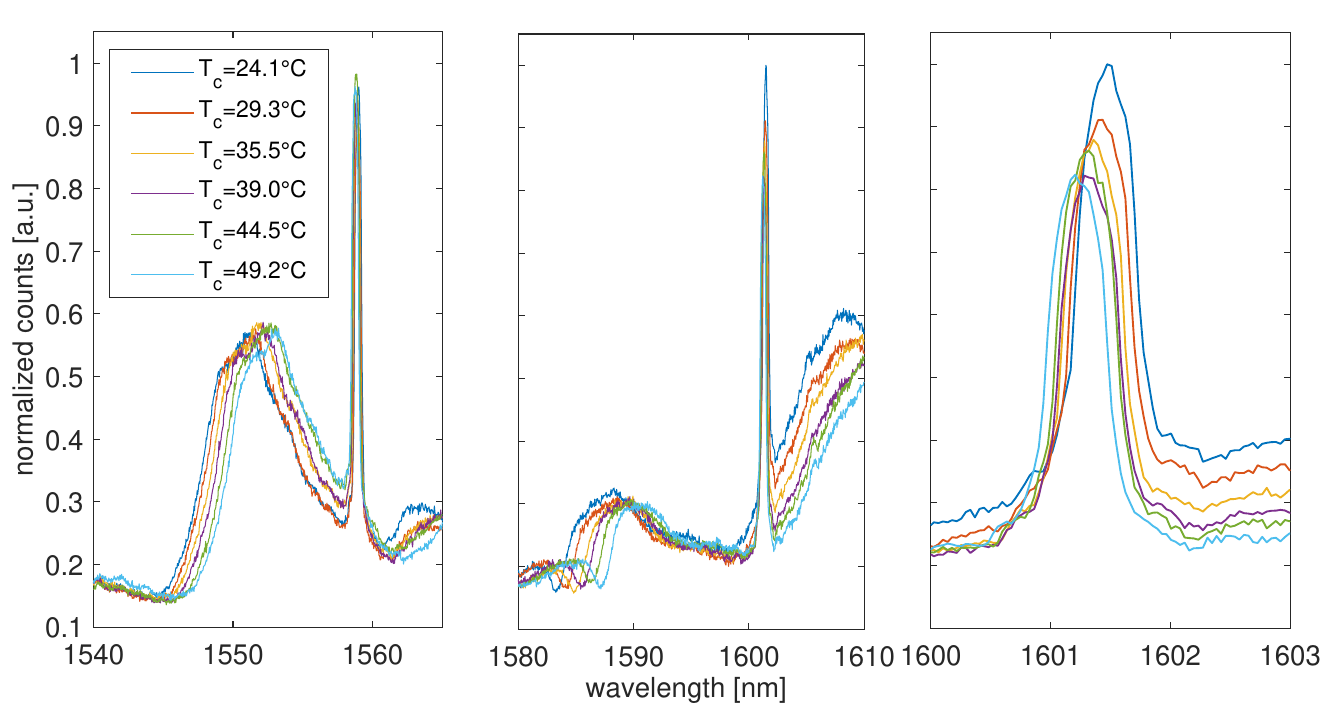}
\caption{Examples of crystal temperature tuning of spectral features of the noise of the Tailored Photons (TP) crystal from about $25\mbox{ }^{\circ}C$ to $50\mbox{ }^{\circ}C$. As predicted from theory, the counter-propagating SPDC peaks tune slowly to shorter wavelengths for increasing crystal temperature, while the ordinary SPDC noise background tunes considerably faster and in the other direction. The TP crystal data was used here, because it had a stronger continuous background with more distinct noise features than the Raicol crystal. But the same behaviour was observed in the noise of the Raicol crystal (see Table S2 in the supp. doc.).}
\label{fig:temptuning}
\end{figure*}

All narrow peaks (even and odd orders) are substantially broadened up by about a factor of 7-35 -- not only in the region where the idler experiences increased absorption. Compare Table S3 and S4 in the supplemental document. In our preliminary measurements we investigated two crystals with identical specifications from different manufacturers (Raicol and Tailored Photons). The measurements presented here were made with the Raicol crystal, as it was the basis of the converter characterisation of Ref. \cite{mann2} and thus provides an accurate normalisation for the NSD. Notably, while the two crystals showed qualitatively similar noise spectra, there were a few distinct differences. The process where the signal counter-propagates is only evident in measurements of the Raicol crystal. We attribute this to the Tailored Photons (TP) crystal having an approximately 5 times larger background for the ordinary noise process when compared to the Raicol crystal, effectively burying the higher-order process. This suggests the random duty cycle errors in the TP crystal are larger than those present in the Raicol crystal. However, despite this reduced background, the peaks of the counter-propagating processes evident in the spectrum taken with the Raicol crystal are considerably broader than those of the TP crystal. This suggests the Raicol crystal might have comparatively larger poling length errors, that, while evident in this measurement, are not deleterious to the SPDC noise floor.

Our interpretation of the measured background is further supported by measurements of the temperature dependence of the NSD. As shown in Fig.~\ref{fig:temptuning}, the broad SPDC background demonstrates a clear temperature dependence expected from a co-propagating process. In contrast, the narrow peaks demonstrate the comparatively weak temperature dependence expected from counter-propagating processes. 

Knowledge of these counter-propagating noise processes might provide valuable insight into the underlying quasi-phase-matching, or the crystal properties. For example, with good knowledge of the poling period, and the Sellmeier equations for the signal wavelength, the location of the peaks at the signal wavelength allows inference of the Sellmeier equations at the idler wavelength. Alternatively, the periodic nature of these features might also allow for improved precision in inferring the grating period. Furthermore, the appearance or absence of phase-matching associated higher-order polings may complement our understanding of poling errors, especially when considered alongside the size of the homogeneous-background attributed to random duty cycle errors.

\section{Theory of higher-order counter-propagating SPDC}
   \label{chap:counterpropagation}

The efficiency $\eta_{spdc}$ of the m-th order quasi-phase-matched SPDC process is given by \cite{fejer,phillips,fiorentino,schneeloch}
\begin{gather}
    \eta_{spdc} \propto \left| E_p \cdot L \cdot d \cdot \frac{2}{\pi m} \cdot \sin{(\pi m D)}\cdot \mbox{sinc}\left(\frac{\Delta k_m^{\mp \mp} L}{2}\right)\right|^2.
    \label{formula:spdc}
\end{gather}
Here, $E_p$ is the pump field, L the length of the crystal, $d$ the nonlinear coefficient, and D the duty cycle of the periodic poling. Note that the efficiency of higher-order processes scales with order $m$ as $1/m^{2}$. Thus, for large $m$ the efficiency decreases relatively slowly with increasing order $m$. Due to the $\sin(\pi m D)$ term in the above equation, even for small deviations of D from 50:50 (e.g. D=0.49), both odd and even orders of large m (e.g. m>10) can be observed in the spectrum. The phase-mismatch of a SPDC process, taking into account $m$-th order quasi-phase-matching, is given by
\begin{gather}
    \Delta k_m^{\mp \mp} = \Delta k_{spdc}^{\mp \mp}- m\Delta k_g\hspace{0.5cm}\mbox{with}\hspace{0.5cm} m \in \mathbb{Z}.
\end{gather}
Here the grating vector $\Delta k_g$ is given by the poling period $\Lambda$ via $\Delta k_g = \frac{2\pi}{\Lambda}$. The phase-mismatch $\Delta k_{SPDC}^{\mp,\mp}$ for all four possible down-conversion processes of the pump is given by
\begin{gather}
    \Delta k_{spdc}^{\mp,\mp}= k_p \mp k_s \mp k_i\hspace{0.5cm}\mbox{with}\hspace{0.5cm} k_x=\frac{2\pi n(\lambda_x)}{\lambda_x},
\end{gather}
with the subscript ether denoting pump (p), signal (s) or idler (i). The "-/+"-sign indicates that the generated photon is travelling in the same/opposite direction as the pump photon. For ordinary/counter-propagating processes the signs for signal and idler are equal/different.

Further, the detuning characteristics of $\Delta k_{spdc}^{\mp,\mp}$ for changes in signal frequency $\omega_s$ and crystal temperature $T$ are important to understand the measured data in Section \ref{chap:parametric}. For changing the signal frequency $\omega_s$ by $\delta \omega$ we have
\begin{equation}
\begin{gathered}
    \Delta k_m^{\mp,\mp}(\Tilde{\omega}_s=\omega_s+\delta\omega) = \frac{1}{c_0}\left[\omega_p n_p\mp(\omega_s+\delta\omega) n_s\mp(\omega_i-\delta\omega) n_i \right]-m\Delta k_g\\
    = \Delta k^{\mp,\mp}_m(\omega_s) + \delta\omega \cdot \frac{(\mp n_s \pm  n_i)}{c_0}.
\end{gathered}
\end{equation}
Here $\omega_p=\omega_s+\omega_i$ was used. Thus, counter-propagating processes detune very fast in signal frequency compared to the ordinary processes. This also leads to different bandwidths in frequency - these highly non-degenerate processes are approximated in first order by 
\begin{gather}
    \Delta \nu^{\mp} = \frac{c_0}{2\pi L}\frac{2 \cdot 2.7831}{(n_{g,s}\mp n_{g,i})}.
\end{gather}
Here $n_{g,x}$ are the group indices. The "-"-sign is for ordinary and the "+"-sign for counter-propagating processes. Note that the counter-propagating processes are about 100 times more narrow than the ordinary processes and are thus apparent as very narrow peaks in the spectrum. The bandwidths of the relevant processes are that of the converter: 109 GHz (0.92 nm at 1587 nm), the ordinary processes: 354 GHz (2.97 nm at 1587 nm) and the counter-propagating processes: 3.6 GHz (30 pm at 1587 nm). When detuning the crystal temperature $T$ by $\delta T$ we have
\begin{equation}
\begin{gathered}
    \Delta k_m^{\mp,\mp}(T+\delta T) = \frac{1}{c_0}\left[\omega_p n_p(T+\delta T)\mp\omega_s n_s(T+\delta T)\mp\omega_i n_i(T+\delta T) \right] - m  \Delta k_g(T+\delta T)\\
    \approx \Delta k_m^{\mp,\mp}(T) + \frac{T}{c_0}\left[\omega_p \partial_T n_p\mp\omega_s \partial_T n_s\mp\omega_i \partial_T n_i \right] - m T \partial_T\Delta k_g.
\end{gathered}
\end{equation}
Thus, ordinary processes detune fast with temperature and counter-propagating processes detune slowly. For counter-propagating processes, a substantial contribution of the temperature tuning (about half) is due to the temperature dependence of the grating vector $\Delta k_g$. The temperature dependence of the grating vector is due to the thermal expansion of the poled domains. Since the grating vector is multiplied by m, for bigger m, this contribution increases.
\\Further, at a peak $\Delta k_m$ becomes zero, so at the peaks we have
\begin{equation}
\begin{gathered}
    \Delta k_m \overset{!}{=} 0 \\
    \Delta k_{spdc}^{\mp,\mp}- m\Delta k_g \overset{!}{=} 0 \\
    \Rightarrow m=\frac{\Delta k_{spdc}^{\mp,\mp}}{\Delta k_g}.
\end{gathered}
\end{equation}
Thus the fraction $\frac{\Delta k_{spdc}}{\Delta k_g}$ equals the phase-matching order m at the peaks (compare the four colored axes in Fig.~\ref{fig:counterpropagation1}). These fractions are plotted as additional axes above Fig.~\ref{fig:counterpropagation1}. From the distance of the peaks relative to each other, the grating period $\Lambda$ can be calculated
\begin{equation}
\begin{gathered}
    \Delta k_{m+1}^{\mp,\mp}-\Delta k_m^{\mp,\mp}\overset{!}{=}0\\
     \left(\Delta k_{spdc}^{\mp,\mp}- (m+1)\Delta k_g\right)-\left(\Delta\Tilde{k}_{spdc}^{\mp,\mp}-m \Delta k_g\right)\overset{!}{=}0\\ 
     \Delta k_{spdc}^{\mp,\mp}-\Delta\Tilde{k}_{spdc}^{\mp,\mp}-\Delta k_g\overset{!}{=}0\\
    \Rightarrow \Lambda=\frac{2 \pi}{\Delta k_{spdc}^{\mp,\mp}-\Delta\Tilde{k}_{spdc}^{\mp,\mp}}.
\end{gathered}
\end{equation}
This could be used to calibrate Sellmeier equations and effective poling periods, e.g. already a 0.1\% different poling length would shift the sharp peaks measurably. 

\section{Conclusion}
In this work we have provided a full characterisation of the noise spectrum from 1140 nm to 1650 nm of a bulk ppKTP quantum frequency converter pumped at 1064 nm, making use of measurements over a very broad spectral range combining both a large dynamic range and high spectral resolution. Through selection of the polarisation of the pump and detected light, as well as tuning of the crystal temperature, different peaks associated with Stokes-Raman scattering were identified in the range from 1140 nm to 1330 nm  (up to $\approx$ 60 THz from the pump) by comparing the noise spectrum to previous Raman scattering measurements in KTP \cite{kugel,ramanKTP1,ramanKTP2}. At photon energies $\approx$ 60 THz or lower than that of the pump light -- between 1330 nm to 1650 nm -- the noise was found to be dominated by broadband parasitic SPDC and phase-matched high-order counter-propagating SPDC processes. 

These higher-order counter-propagating SPDC processes were observed in the spectrum in the form of a characteristic succession of very narrow-band peaks of different heights. We believe this phenomenon was heretofore unobserved experimentally. Additionally, phase-matching of even orders of m, enabled by small poling errors related to the duty cycle was also visible.

As these counter-propagating SPDC processes are very sensitive to the material dispersion of the host nonlinear crystal, they may find application in refining Sellmeier equations or the precise measurement of gratings. The enhanced sensitivity of these counter-propagating SPDC processes to poling errors also suggests they could be exploited to provide insights into these errors and their distribution, e.g. through the existence or absence, the relative brightness and the widths of peaks. Our presented measurement techniques - in essence combining high power pumping at 1064 nm and measurements with off-the shelf InGaAs-grating spectrometers - both of which are technologically rather straight-forward and moderately cost-effective - might offer another approach to better understand and design improved periodically poled nonlinear devices.

We expect that the results of this in-depth investigation of the noise spectrum of a bulk ppKTP quantum frequency converter will be important for the further development of frequency converters for future quantum networks and quasi-phase-matched devices for quantum technology applications in general.

\begin{backmatter}

\bmsection{Funding Information} Funded by the BMBF, Germany within the project QR.X.

\bmsection{Acknowledgments} We thank William Staunton and Oliver Benson for providing equipment.

\bmsection{Disclosures} The authors declare no conflicts of interest.

\bmsection{Data Availability Statement} Data underlying the results presented in this paper are not publicly available at this time but may be obtained from the authors upon reasonable request.
\newpage

\bmsection{Supplemental document}
\renewcommand{\thefigure}{S1}
\begin{figure*}[!htbp]
\centering
\includegraphics[width=0.45\textwidth]{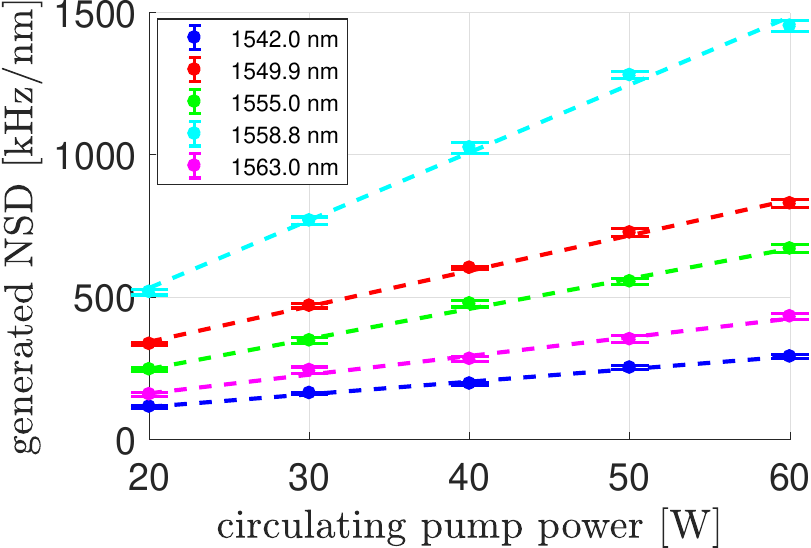}\quad
\includegraphics[width=0.45\textwidth]{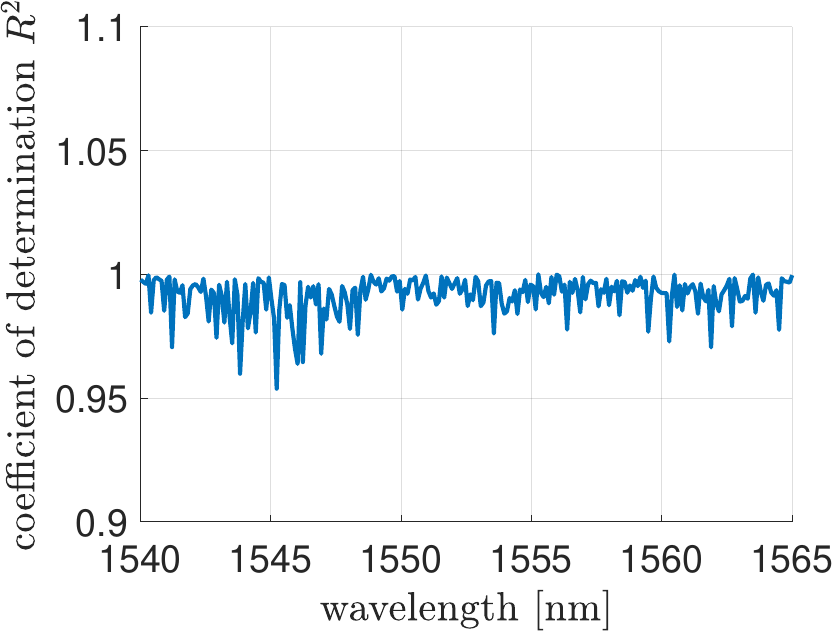}\vspace{10px}
\includegraphics[width=0.45\textwidth]{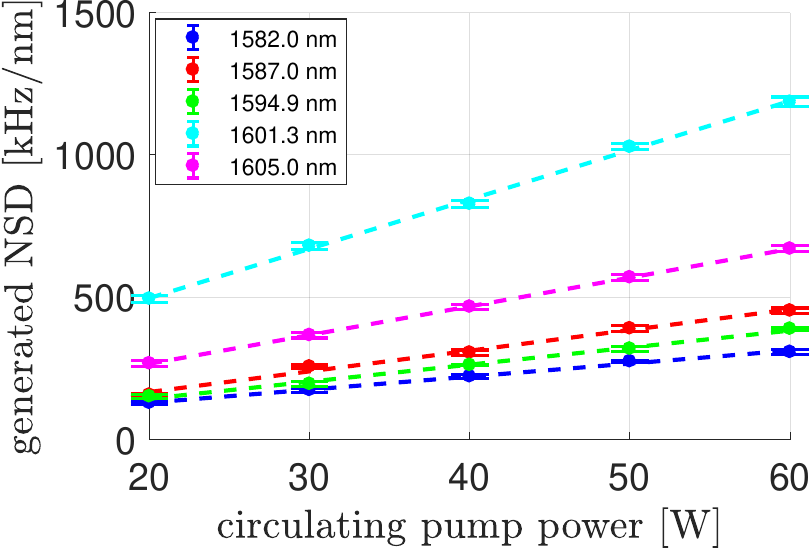}\quad
\includegraphics[width=0.45\textwidth]{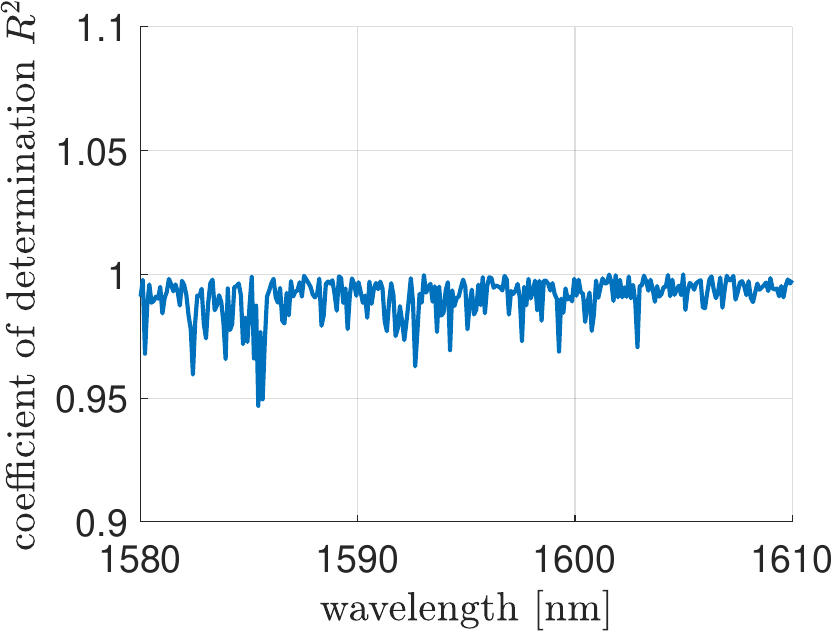}
\caption{Pump laser power scaling for the noise spectral density of the Tailored Photons (TP) crystal at $35.5\mbox{ }^{\circ}C$. The first row corresponds to the spectral region of $1540 - 1565\mbox{ nm}$, and the second row to $1580 - 1610\mbox{ nm}$. The first column shows the linear power scaling behavior for several wavelengths, chosen to represent different spectral features (refer to Figure 7 in the main document). The second column presents the coefficient of determination ($R^2$) of the linear fit for each wavelength.}
\label{fig:spdcpowerscaling}
\end{figure*}

\begin{table}[!htbp]
\renewcommand\thetable{S1}
\centering \caption{Measured Raman resonances compared to the results from Kugel et al. \cite{kugel} (compare groups VII, VIII, IX in Table 1 and groups VIII and IX in Table 3 in the reference).}
\begin{tabular}{ccc}
\hline \hline
$\lambda_{peak}$ [nm] & $\Delta \tilde{\nu}_{peak}$ [1/cm] & $\Delta \tilde{\nu}_{peak}$ [1/cm] (Ref. \cite{kugel})\\ \hline
$1149.4\pm 0.8$ & $692\pm 6$ & 687.5\\ 
$1167.2\pm 0.3$ & $825\pm 2$ & 826.2\\
$1188.5\pm 0.7$ & $979\pm 5$ & 982\\
$1196.2\pm 2.8$ & $1033\pm 20$ & 1028.9\\
$1207.9\pm 0.9$ & $1114\pm 6$ & 1113\\ \hline
$1240.8\pm 0.6$ & $1333.3\pm 4$ & -\\ \hline
$1277.9\pm 4.6$ & $1567\pm 28$& -\\
$1290.6\pm 8.3$ & $1644\pm 50$ & -\\
$1319.0\pm 1.4$ & $1811\pm 8$ & -\\ \hline \hline
\end{tabular}
\label{table:raman}
\end{table}

\begin{table}[!htbp]
\renewcommand\thetable{S2}
\centering \caption{Temperature tuning of the different processes for the Raicol (R) and the Tailored Photons (TP) crystal. Compare Equation 3 in the main document for the labelling of the processes with "$\mp \mp$".}
\begin{tabular}{cccccc}
\hline \hline
crystal & process & $\lambda$ [nm] & meas. $d\lambda/dT$ [pm/K] & pred. $d\lambda/dT$ [pm/K]& deviation\\ \hline
 & -+ & 1558 & -9 & -10 & 10\% \\ 
Raicol & +- & 1565 & +20 & +23 & 13\% \\ 
 & - - & 1580 & +80 to +130 & +269 & x2-x3 \\ \hline 
 & -+ & 1559 & -11 & -10 & 10\% \\
TP & - - & 1585 & +155 & +279 & x2 \\
 & - - & 1554 & +76 & +219 & x3 \\ \hline \hline
\end{tabular}
\label{table:temp}
\end{table}

\begin{table}[!htbp]
\renewcommand\thetable{S3}
\centering \caption{Center wavelengths and bandwidths (FWHM) of the high counter-propagating peaks from the Raicol (R) and the Tailored Photons (TP) crystal. Compare Equation 3 in the main document for the labelling of the processes with "$\mp \mp$".}
\begin{tabular}{ccccc}
\hline \hline
order & 15th & 16th & 17th & 18th \\ \hline
process & -+ & -+ & -+ & -+ \\ \hline
$\lambda_{center,R}$  [nm] & 1480.10 & 1518.13 & 1558.16 & 1600.59 \\ 
$\Delta\lambda_{meas,R}$ [pm] & - & 495 & 529 & 516 \\
$\Delta\lambda_{theory,R}$ [pm] & 26 & 27 & 29 & 30 \\ \hline
$\lambda_{center,TP}$ [nm] & 1480.60 & 1518.73 & 1558.86 & 1601.39 \\
$\Delta\lambda_{meas,TP}$ [pm] & 212 & 185 & 193 & 195 \\
$\Delta\lambda_{theory,TP}$ [pm] & 26 & 27 & 29 & 30 \\ \hline \hline
\end{tabular}
\label{table:peakwidth1}
\end{table}

\begin{table}[!htbp]
\renewcommand\thetable{S4}
\centering \caption{Center wavelengths and bandwidths (FWHM) in terms of wavelengths of the small counter-propagating peaks from the Raicol crystal (R). Compare Equation 3 in the main document for the labelling of the processes with "$\mp \mp$".}
\begin{tabular}{ccccc}
\hline \hline
order & 36th & 37th & 38th & 39th \\ \hline
process & +- & +- & +- & +- \\ \hline
$\lambda_{center,R}$ [nm] & 1607.79 & 1565.06 & 1524.63 & 1486.40 \\
$\Delta\lambda_{meas,R}$ [pm] & 712 & 979 & 840 & 904 \\ 
$\Delta\lambda_{theory,R}$ [pm] & 31 & 29 & 27 & 26 \\ \hline \hline
\end{tabular}
\label{table:peakwidth2}
\end{table}
\newpage
\end{backmatter}
\newpage



\end{document}